\documentstyle[prl,aps,epsf]{revtex}   
\begin{document}
\twocolumn[\hsize\textwidth\columnwidth\hsize\csname %
@twocolumnfalse\endcsname

\title{Low Temperature Magnetic Properties of the Double Exchange Model}
\author{Jun Zang, H. R\"oder, A.R. Bishop, S.A. Trugman}
\address{
    Theoretical Division and Center for Nonlinear Studies, MS B262
    Los Alamos National Laboratory, \\
    Los Alamos, NM 87545}
\date{Received: May 2 1996}
\maketitle
\begin{abstract}
We study the {\it ferromagnetic} (FM) Kondo lattice model
in the strong coupling limit (double exchange (DE) model). 
The DE mechanism proposed by Zener to explain ferromagnetism
has unexpected properties when there is more than one
itinerant electron.
We find that, in general,
the many-body ground state of the
DE model is {\it not} globally
FM ordered (except for special filled-shell cases). 
Also, the low energy excitations
of this model are distinct from spin wave excitations in 
usual Heisenberg ferromagnets,
which will result in unusual dynamic magnetic properties.
\end{abstract}
\phantom{.}

]

\narrowtext
\pagebreak

The double exchange (DE) model  \cite{Zen51,And55,deG60} 
has attracted much recent  attention
\cite{Kub72,wang,furukawa,Mil95,jp95,hzb1,hzb2,hartmann} because of
its anticipated relation to the Mn-oxide perovskites 
$La_{1-x}A_xMnO_3$ (${A}=Ca,Sr,Ba$) materials with 
colossal magnetoresistance (CMR) \cite{jonker,volger,wollan,von93,jin94}.
To explain ferromagnetism in these Mn-oxides,
Zener \cite{Zen51} introduced a
DE mechanism, in which local $S=3/2$ spins of the
three Mn $t_{2g}$ $d$-electrons become ferromagnetically coupled due to
coherent hopping of $e_g$ electrons. 

The DE mechanism can be derived from the following
{\it ferromagnetic} (FM) Kondo lattice model
Hamiltonian \cite{And55,Kub72}
\begin{equation}
  H = - t\sum_{\langle i,j\rangle\alpha} 
  c_{i\alpha}^{\dagger}c_{j\alpha}^{\phantom{\dagger}} 
- J_H\sum_{i} \vec{S}_i\cdot\vec{\sigma}_{\alpha\beta}^{\phantom{\dagger}} 
c_{i\alpha}^{\dagger}c^{\phantom{\dagger}}_{i\beta},
\label{eq:hde0}
\end{equation}
where $\vec{\sigma}$ is the Pauli matrix and $\vec{S}_i$ is the
local spin $S=3/2$ of three Mn $t_{2g}$ $d$-electrons. The operators 
$c_{i\alpha}^{\phantom{\dagger}}$ 
($c_{i\alpha}^{\dagger}$) annihilate (create) a 
mobile $e_g$ electron with spin $\alpha$ at
site $i$. The DE mechanism requires the Hund's rule
coupling $J_H\gg t/S$. 

The Hamiltonian (\ref{eq:hde0}) was studied by Anderson
and Hasegawa \cite{And55} using a single electron on two sites.
They concluded that, in the low-energy band,
the electron spin must be parallel to the local spin at every site,
leading to a renormalization of the electron hopping integral
by $\langle S_0^{ij} +\frac{1}{2}\rangle / (2 S +1)$,
where $S_0^{ij}$ is the total spin of the subsystem consisting
of the two localized spins on sites $i$ and $j$ and the electron.
Based on this single electron result, 
in most theoretical calculations the original model (\ref{eq:hde0})
is mapped onto a non-interacting model of spinless fermions with spin disorder
scattering \cite{Mil95,hzb1}. 
In this non-interacting model, the low temperature
phase is ferromagnetic (FM), due to the gain of electron kinetic energy
if the local spins are aligned. This DE mechanism is also implemented
in many-body treatments\cite{Kub72,furukawa,Mil95,hzb2} 
by using FM mean-field solutions or by expanding around them.
However, in the original model (\ref{eq:hde0})
the ground states are {\it not} FM for some fillings due to 
the Pauli principle. This is readily seen in the half-filling
case (see below), in which the ground state
has {\it antiferromagnetic} (AFM) correlations.
In this Letter, we study the Hamiltonian (\ref{eq:hde0}) using
exact diagonalization and variational wavefunctions.
We show that the DE mechanism is a purely
single electron property, and that the 
many-body ground states are typically {\it not} FM \cite{note},
except for special fillings in which the non-interacting states
at the Fermi level are completely filled.
 The ground states
in model (\ref{eq:hde0}) are generally 
only {\it locally} FM without
global FM ordering.
Also, the low energy excitations are different from spin wave excitations in
Heisenberg FM systems even if the ground state has FM ordering
at special fillings. 

The Hamiltonian (\ref{eq:hde0}) conserves total momentum $\vec{K}$,
total spin $S^t$, and $S^t_z$. For the FM states,
$S^t=S^t_{max}=NS+n/2$, where $N$ is the number of sites, and $n$ is the
number of electrons. It can be easily seen that the wavefunctions
of low energy FM eigenstates are simply the Slater determinants of
spinless fermions
$|\vec{k}_1,\vec{k}_2,\cdots,\vec{k}_n\rangle\otimes|FM\rangle$, with
energy $\sum_i(-J_HS+\epsilon_{\vec{k}_i})$, where $\epsilon_{\vec{k}}$
is the non-interacting electron dispersion.
$|FM\rangle$ denotes the FM states for local spins, with the
electronic spins being fixed by local spins.
We will denote the
lowest FM eigenenergy by $E_0$.
For our numerical calculations,
we set $t=1$, $J_H=20$.
In the manganese perovskites $S=3/2$, however
we will take $S=1/2$ for most of our numerical 
calculations; some calculations
are repeated for smaller systems at $S=3/2$, and
we obtain qualitatively the same results. 

For one electron $(n=1)$,
it can be shown analytically that in the ground
state $S^t=S^t_{max}$ for $J_H>0$, 
which is consistent with the DE mechanism. At half-filling $n=N$,
there is an induced AFM
coupling $\sim t^2/J_H$ between spins due to the Pauli principle: 
the electrons cannot hop in the FM state;
they can only hop if the neighboring sites are empty or have
anti-parallel spin, inducing an AFM coupling. 
Thus the ground state at half-filling is AFM \cite{ueda}.
The question now is, what happens for general fillings.
We can take some guidance from a heuristic 
non-interacting electron picture: 
the electrons have to be parallel to the local spin at {\it every}
site, since  $J_HS/t\gg 1$.  Due to the 
renormalization of the electron hopping integral 
$t\rightarrow
t \langle S_0^{ij} +\frac{1}{2}\rangle / (2 S +1)$,
mobile electrons have the largest bandwidth
in the FM states; electronic
spins in these states are totally polarized, and then fill states up to
the Fermi level. Non-FM states of the local spin
will reduce the bandwidth; but the electronic spins are
not totally polarized, and so can have a lower 
kinetic energy.
Thus it is plausible that
at some fillings the ground state can be non-FM.
It is straightforward to see that non-ferromagnetic
ground states exist in the limit where the
local spins are classical $ ( S \rightarrow \infty )$.
Consider a 1D ring where the local spins uniformly circle
the north pole at an angle $\theta$.
The magnitude of the effective
hopping $t_{eff}$ is reduced as $\theta$ increases, but an
additional magnetic flux (Berry phase) is simultaneously
added.  It is easy to verify that for large $J_H$ the change in energy
$\Delta E = - c_1 t^2 \sin ^2 (\theta ) / J_H $,
where $c_1$ is a positive constant for an even number of electrons.
For finite $J_H$, the Berry phase reduces the total energy
by more than $t_{eff}$ increases it, and the
spins prefer to lie on the equator \cite{xxx}.

That this mechanism is also relevant for small $S$ can be
shown for two electrons in any dimension, by
constructing a wavefunction that is not FM but lower in energy than
$E_0$.
We denote the zone center $\vec{k}_0=0$ with (lowest Bloch) 
energy $\epsilon_0$, and the first excited states with energy $\epsilon_1$
by $\vec{Q}_i$, with $i=1,\cdots,M$
 ($M\ge 2$). Then we can write a wavefunction
with $S^t=S^t_{max}-1$ as follows: 
\begin{eqnarray}
|\Psi\rangle = \sum_{i=1}^{M}\sum_l e^{iQ_il}(S^-_l
+xc^{\dagger}_{l\downarrow}
c_{l\uparrow})^{\phantom{\dagger}}\,c^{\dagger}_{0\uparrow}
c^{\dagger}_{-Q_i\uparrow}
\,|0\rangle |FM\rangle ,
\label{eq:wf1}
\end{eqnarray}
where $|0\rangle$
is the electronic vacuum state. The variational energy $E_v$
decreases with increasing $J_H$.
As $J_H/t\rightarrow \infty$, $x\rightarrow 1$, and
$E_v=E_0-(\epsilon_1-\epsilon_0)(M-1)/(2SN+M+1)$.
If the number of sites $N>2$, the energy difference $E_0-E_v$
is finite as $J_H\rightarrow \infty$.
Our exact diagonalization calculation shows that
the ground state is $S^t=0$ in 1D and 2D for 
two electrons $(n=2)$. 
Consequently, the trial wavefunctions can only tell us that
there exist lower energy non-FM states.
These trial wavefunctions have
finite probability of {\it two} electrons being
in $|\vec{k}_0=0\rangle$
by using the degeneracy
of the non-interacting states at the Fermi level. 

In 1D, we can extend the results for $n=2$ to general fillings:
The ground states for even
numbers of electrons are non-FM for any $n=2m$ ($m$ integer)
and $N$.
This can be proved by constructing the 
following wavefunction with
$S^t=S^t_{max}-1$\cite{note-wf}:
\begin{eqnarray}
|\Psi\rangle &=&\sum_l e^{-iq_0l}(S^-_l+xc^{\dagger}_{l\downarrow}
c^{\phantom{\dagger}}_{l\uparrow})\,|vac\rangle -{x\over 2NS+2m}
\nonumber \\
 & &\sum_l (S^-_l+c^{\dagger}_{l\downarrow}
c^{\phantom{\dagger}}_{l\uparrow})\,c^{\dagger}_{-mq_0\uparrow}
c^{\phantom{\dagger}}_{(1-m)q_0\uparrow}\,|vac\rangle ,
\label{eq:wf2}
\end{eqnarray}
with  $|vac\rangle=c^{\dagger}_{(1-m)q_0\uparrow}
\cdots c^{\dagger}_{mq_0\uparrow}|0\rangle |FM\rangle$ and $q_0=\pm 2\pi/N$. 
As for $n=2$, the variational energy $E_v$ 
decreases with increasing $J_H$.
It can be shown that, for $J_H/t\rightarrow \infty$, $x\rightarrow 1$, and
$E_v<E_0$. So there is at least one eigenstate with $S^t=S^t_{max}-1$,
$K=(m-1)q_0$ with energy  less than $E_0$.
We believe that the difference between even and odd 
number of electrons arises as follows: In the FM states, the
electrons fill the non-interacting states up to the Fermi level;
for an odd number of electrons, the $\pm k_f$ states at the Fermi level are
all filled; for an even number of electrons, only one of the $\pm k_f$
states is filled. This kind of degeneracy is related to
the possibility of having non-FM ground states. The wavefunctions
(\ref{eq:wf1}) and (\ref{eq:wf2})
have a lower energy than $E_0$ due to this degeneracy. 
In 1D systems, we calculated the ground state and some low energy excited
states using Lanczos methods for sizes up to $N=16$, $n=2$; and
$N=10$, $n=6$. 
We find that the ground state 
is always $S^t=0$ and $K=\pi$ for even numbers
of electrons, and
FM and $K=0$ for odd numbers of electrons. 

\begin{figure}
 \centerline{
 \epsfxsize=7.5cm \epsfbox{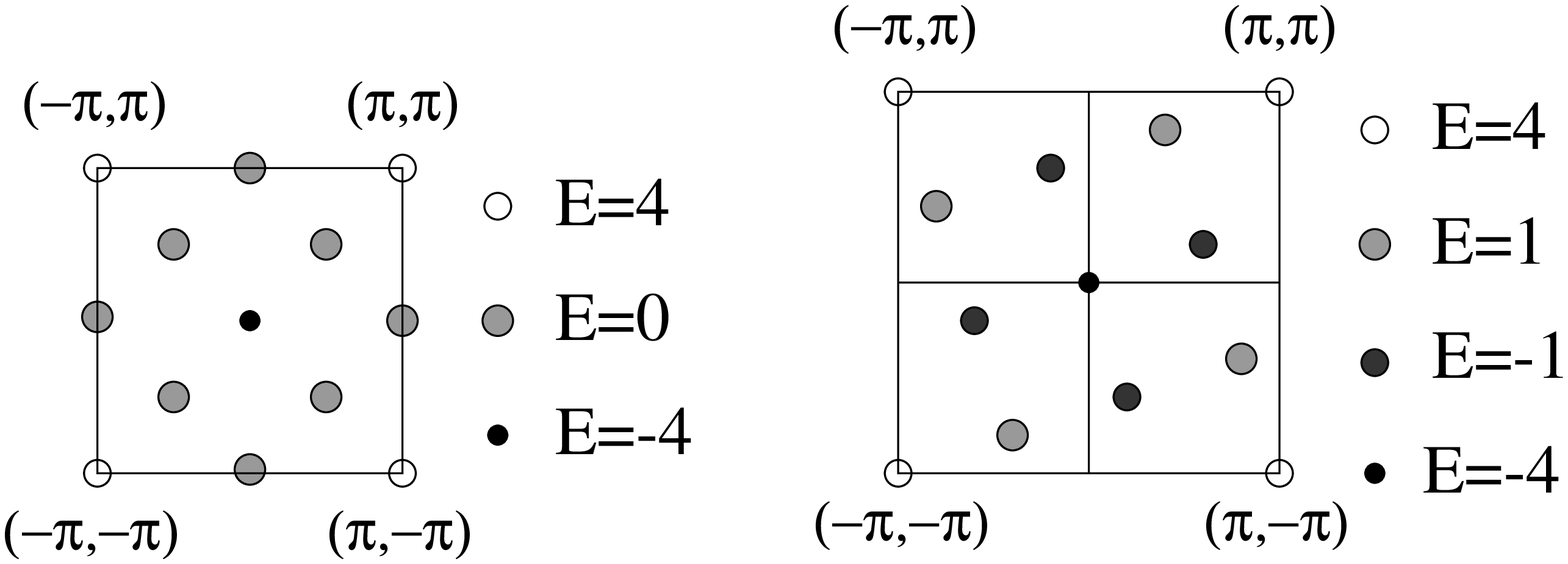}
}
\caption{ Brillouin zone of the 2D $N=8$
  and $N=10$ lattice. }
 \label{fig:bz} 
\end{figure}

\begin{figure}
 \centerline{
 \epsfxsize=2.5cm \epsfbox{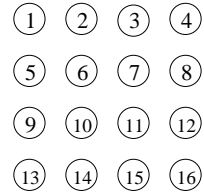}
}
\caption{Site indices for the 2D $4\times 4$ lattice.  \label{fig:lat} }
\end{figure}

The trial wavefunction in 1D cannot be generalized
to higher dimensions except for $n=2$. However, we believe
that the origin
of the non-FM ground state should be the same for any dimension. 
Recall that our results in both the $n=N$ and the $n=2$ cases are
independent of dimensionality. 
We speculate that the
ground state of  model (\ref{eq:hde0}) is non-FM in higher dimensions
whenever the single-particle states at the Fermi level of the FM ground state
are not completely filled. 
This has been confirmed in our finite size
exact diagonalization calculations.
The filling of the non-interacting single
electron states plays an important role here
because the wavefunction of the FM ground state is just a Slater determinant
wavefunction of the spin polarized non-interacting electron system.
From the single electron dispersion of the non-interacting 
systems (Fig.(\ref{fig:bz})), we see that the filled-shell cases
are 1, 7 for the $\sqrt{8}\times\sqrt{8}$ lattice
and 1, 5, 9 for the $\sqrt{10}\times\sqrt{10}$ lattice.
For these fillings, we find that the ground states are FM.
The non-filled-shell ground states have 
\newline
$S=0\,(2),1/2\,(3),0\,(4),1/2\,(5),0\,(6),0\,(8)$,
for N=8, \newline
$S=0\,(2),5/2\,(3),4\,(4),7\,(6)$,
for N=10 and \newline
$S=0\,(2)$ for N=16.
\newline
(The number in parenthesis is the number of electrons).
 In all these cases the ground states are non-FM.
We also calculated the $S=3/2$ 
case up to $N=8$ and $n=4$. The results are similar to $S=1/2$. 
One different feature in a 2D (e.g. $\sqrt{10}\times\sqrt{10}$) lattice
is that the ground
states are not necessarily at the smallest $S^t$ 
for non-filled shell fillings. But for all the cases we have checked,
there is at least one state in each $S^t<S^t_{max}$ 
which has a lower energy than $E_0$.

\begin{figure}
 \centerline{
\epsfxsize=7.5cm \epsfbox{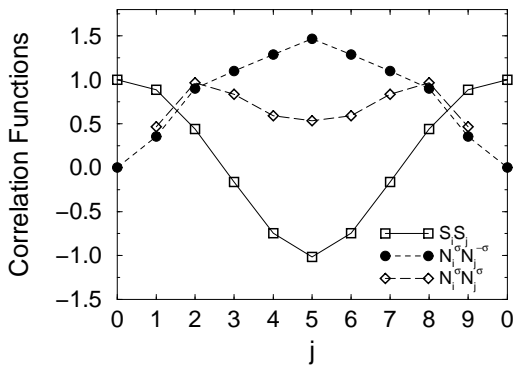}
}
\vskip -10pt
\caption{Correlation functions $\langle n_1^{\sigma} n_{j+1}^{-\sigma}\rangle$, 
$\langle n_1^{\sigma} n_{j+1}^{\sigma}\rangle$,
$\langle \vec{S}_1\cdot\vec{S}_{j+1}\rangle$ 
for the 1D $N=10$, $n=4$, $S=1/2$ system.
The curves are rescaled to fit on the same graph, 
so only the relative magnitudes
are representative.
\label{fig:corr} }
\end{figure}

What is the magnetic structure of the non-FM ground state?
At $n=N$, the ground state is AFM as described above.
At $n=2$, the ground state 
also has $S^t=0$, but the spins have {\it local} FM
ordering, which is evident
in the correlation functions (see Table \ref{tab:str} for a
$4\times4$ 2D lattice).
For other fillings
with non-FM ground states $S^t\ll S^t_{max}$, the magnetic
structures are most likely to be locally but not globally
FM ordered (see 
Fig.(\ref{fig:corr})).
For 1D systems, the local spin structures are  
{\it spiral} with a pitch equal
to the length of the chain. This can be seen in the correlation
function 
$\langle(\vec{S}_1\times\vec{S}_2)\cdot(\vec{S}_i\times\vec{S}_{i+1})\rangle$,
which is only weakly dependent on $i$ and has magnitude
$\sim (0.25\sin(2\pi/N))^2$. 
Preliminary numerical results suggest 
that the 2D non-FM ground states are also noncolinear
for non-filled-shell cases. 
The details of the spin structures for 1D and higher
D are not the major concern here
and will be discussed elsewhere.
What we can conclude from these correlation functions 
(see Fig.(\ref{fig:corr}) and Table \ref{tab:str}) is that
the spin-up electrons are repelled from spin-down electrons and
the local spins with greatest separation are anti-parallel: thus there is
local (i.e. FM domains), but not global FM ordering.

\begin{figure}
\vskip -10pt
\centerline{
\epsfxsize=7.0cm \epsfbox{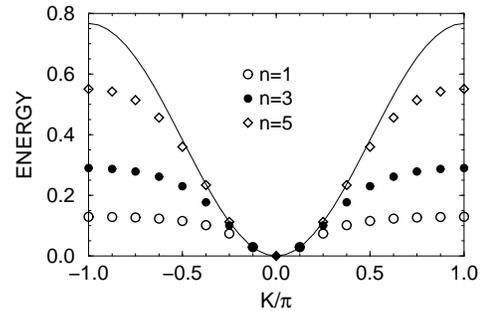}
}
\vskip -10pt
\caption{ Dispersion of SW states at $N=16$,
$n=1,3,5$. We have rescaled the $n=1,3$ curves.
The solid line is the cosine
curve fitted by the lowest two points.
Parameters: $S=1/2$, $t=1$, $J_H=20$. \label{fig:sw} }
\end{figure}

The excitation spectrum of model (\ref{eq:hde0}) is also unusual.
For the filled-shell cases, there are quasi ``spin wave'' (SW)
excitations. 
However, there are many other branches of $\Delta S^t=1$ excitations
in model (\ref{eq:hde0}). For the quasi SW states,
the dispersion is not a cosine, as it would
be in a near-neighbor Heisenberg ferromagnet (see Fig.(\ref{fig:sw})).
In Fig.(\ref{fig:sw})
we show the dispersion of SW states
of a 1D system with $N=16$, $n=1,3,5$.
From Fig.(\ref{fig:sw}), we can see that the effective
spin-spin coupling $J$ softens at short wavelength, 
and the softening decreases with
increasing electron density.
This softening at short wavelength (large wavenumber)
is due to a spin polaron
effect: the electron density is not uniform and is smallest
at the site with a flipped-spin. For long wavelength SW excitations,
the electron density is nearly constant, so the SW dispersion
is quadratic.  As $K$ increases,
the electrons increasingly avoid the flipped spin, 
so the effective spin-spin
coupling $J$ becomes smaller.
One can show using a Bethe ansatz that the $n=1$
excitations shown in Fig.(\ref{fig:sw})
cannot have an energy greater than
$2t [ 1-\cos ( \pi / N) ]$.  This
is $\frac{1}{4}$ of the lowest electronic
excitation energy, which promotes a free electron
from $K=0$ to $K = 2 \pi / N$. 

\begin{figure}
\vskip -26pt
\centerline{
\epsfxsize=7.0cm \epsfbox{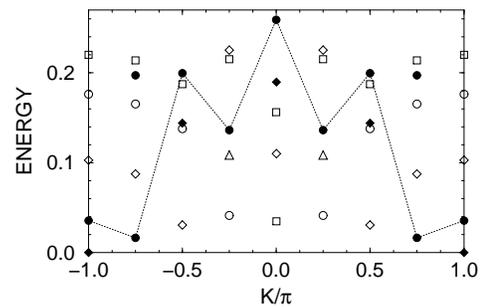}
}
\vskip -10pt
\caption{Dispersion of low energy states at $N=8$,
$n=2$. 
Symbols: filled $\Diamond$: $S=0$; 
$\bullet$: $S=1$; $\Diamond$: $S=2$; 
$\bigcirc$: $S=3$; $\Box$: $S=4$;
$\triangle$: $S=5$. 
Parameters: $S=1/2$, $t=1$, $J_H=20$.   \label{fig:dis} }
\end{figure}

For the non-filled-shell fillings, the low energy
excitation spectrum is complicated (see Fig.(\ref{fig:dis})).
As discussed above, the 1D spin structure in non-filled-shell cases
has spiral ordering. This might suggest that there will also exist quasi SW 
low energy excitations.
However, from Fig.(\ref{fig:dis}), we can
see that it is difficult to interpret the
excitations as spin waves.

We have shown that 
model (\ref{eq:hde0}) exhibits non-FM ground states on finite lattices
for non-filled-shell cases in both 1D and 2D. 
(This has also been confirmed in 3D for up to 4 electrons
on a $3 \times 3 \times 3$ lattice.)
There are many small $S^t$
states with lower energy than the FM ground state energy $E_0$ 
(see Fig.(\ref{fig:dis})).
It is believed that there is weak
AFM super-exchange coupling between local spins
in the CMR materials \cite{deG60,jp95,kawano}. 
The low temperature phase will further deviate
from FM if this is included. 
Experiments on CMR materials show that the low temperature
phase is FM at dopings $\sim 0.2-0.4$ 
with a quadratic long wavelength 
spin wave dispersion at low temperatures \cite{martin,lynn}. 
One possibility is that the model Hamiltonian (\ref{eq:hde0}) does not describe
the physics in the CMR materials:
a more complete model should include, e.g. two interacting orbitals
and electron-phonon (e-p) couplings. 
Another possibility is that the {\it locally} FM ordered phase
found here is experimentally indistinguishable
from global FM ordering. Because of the inhomogeneity 
of electronic density due to disorder or e-p coupling,
the filled-shell phases co-exist with non-filled-shell phases.
Thus there are quasi SW excitations accessible to
 neutron experiments. 
This is of course speculative, but 
 some of the unusual experimental magnetic properties 
\cite{lynn,lofland,bob} in CMR materials
might be explained along this line. 
Notice that the SW excitations of model (\ref{eq:hde0}) in
filled-shell cases is only one of the many branches of $\Delta S=1$ 
excitations; other branches can have considerable higher energies.
Thus $T_c$ in model (\ref{eq:hde0}) may be higher than
estimated using the spin stiffness from SW excitations alone,
i.e. the spin coupling is energy dependent.

In a homogeneous system, the local FM
spiral ordering discussed above might be 
experimentally indistinguishable from the global FM
ordering. However, if a double exchange system
is inhomogeneous and can be separated into small mesoscopic
systems due to Anderson localization, the spiral related
states will have distinct properties from the
FM states. To verify that non-FM states occur
not only for periodic boundary conditions, we have calculated
the ground states for various small clusters using {\it open}
boundary conditions. We find that the ground state
can still be non-FM in $D>1$ finite clusters 
($2\times3$, $3\times4$, etc) with open
boundary conditions.

In this work, we showed that the DE mechanism 
is a single particle mechanism
which does {\it not} give FM in  many-electron systems.
In the DE  model (\ref{eq:hde0}), 
there are many small $S^t$ states with lower energy than the
FM ground state energy $E_0$, except for filled-shell cases. 
In 1D, for an even number of electrons, the
ground state has $S^t=0$ and momentum $K=\pi$. In higher D,
the ground state is FM only if the  non-interacting
states at the Fermi level are 
completely filled. Otherwise there are states 
in each $S^t<S^t_{max}$ subspace,
with energy $<E_0$. 
The spin correlations show that the non-FM
low energy states have local FM-domain-like structures
(spiral ordering in 1D):
two groups of electrons (spin-up, spin-down) 
tend to avoid each other by forming
two ferromagnetic domains. We suggest that 
this low temperature phase might be relevant to the
CMR materials, which have been shown to possess unusual
magnetic properties in recent experiments
 \cite{lynn,lofland,bob}.

{\bf Acknowledgements:} 
We would like to thank E. Abrahams, J.L. Birman, D. Cai, 
T. Egami, R. Heffner, M.F. Hundley, M.B. Salamon,
Z.B. Su, Z. Tesanovic, W.Z. Wang, and C. Yu
for helpful discussions. Work at Los Alamos 
was performed under the auspices of the U.S. DOE, 
and supported (in part) by CULAR funds provided
by the University of California.


\begin{table}
\caption{ The ground state static correlations functions 
for 2D $N=16$ ($4\times4$), $n=2$, $S=1/2$. 
$\langle \vec{S}_i\cdot\vec{S}_j\rangle$ is the correlation
function of the total spin (local + itinerant).
Site indices are shown 
in Fig.(2).
\label{tab:str}}
\begin{tabular}{||c|c|c|c||} 
$(i,j)$ & 
$\langle n_i^{\sigma} n_j^{-\sigma}\rangle$ & 
$\langle n_i^{\sigma} n_j^{\sigma}\rangle$ &
$\langle \vec{S}_i\cdot\vec{S}_j\rangle$ \\
  \hline \hline
(1,2) & 0.003264 & 0.001489 & 0.1655  \\ 
   \hline
(1,3) & 0.005771 &  0.002301 & -0.0131  \\ 
   \hline
(1,6) & 0.005771 & 0.002301 & -0.0131 \\ 
   \hline
(1,7) & 0.008117 & 0.002820 & -0.2333 \\ 
   \hline
(1,11) & 0.010515 & 0.003084 & -0.5559 \\ 
\end{tabular}
\end{table}


\begin{thebibliography}{10}
\bibitem{note} Here we define the non-FM ordering of a finite
system as $S^t < S^t_{max}$.
\bibitem{xxx}  For finite $S$, one can show that
the leading $1/S$ correction is
$\Delta E = - c_1 t^2 \sin ^2 (\theta ) / J_H 
-c_2 |t| \sin ^2 (\theta ) / S$, where $c_1$ and $c_2$ 
are positive constants for an even number of electrons.  
Quantum fluctuations also
favor a single twist state with spins lying
in the equator.
\bibitem{note-wf} Wavefunctions with any $S^Z_{tot}$ can be constructed
similarly.
\end{thebibliography}
\end{document}